\documentclass[aps,prd,onecolumn,amssymb,eqsecnum,nofootinbib]{revtex4}

\usepackage{epsf}  
\usepackage{graphicx}

\begin{document}

\title{Decoherence measure by gravitational wave interferometers}

\author{Yasushi Mino}
\email{mino@tapir.caltech.edu}
\affiliation{mail code 130-33, 
California Institute of Technology, Pasadena, CA 91125}

\begin{abstract} 
We consider the possibility to measure the quantum decoherence 
using gravitational wave interferometers. 
Gravitational wave interferometers create 
the superposition state of photons 
and measure the interference of the photon state. 
If the decoherence occurs, 
the interference of the photon state vanishes 
and it can be measured by the interferometers. 
As examples of decoherence mechanisms, 
we consider 1) decoherence by spontaneous localization, 
2) gravitational decoherence 
and 3) decoherence by extra-dimensional gravity. 
\end{abstract}

\maketitle

%%%%%%%%%%%%%%%%%%%%%%%%%%%%%%%%%%%%%%%%%%%%%%%%%%%%%%%%%%%%
\section{Introduction}
%%%%%%%%%%%%%%%%%%%%%%%%%%%%%%%%%%%%%%%%%%%%%%%%%%%%%%%%%%%%

Gravitational wave interferometers, 
such as LIGO\cite{LIGO}, VIRGO\cite{VIRGO}, 
TAMA\cite{TAMA} and GEO600\cite{GEO}, are 
designed to detect gravitational waves directly. 
By gravitational wave detection, 
we expect to gain better understanding 
of the relativistic theory of gravity. 
They are also expected to serve as a new kind of telescope 
to observe our universe. 
With the anticipation of high sensitivity of interferometers 
in next several years, 
new astronomy using gravitational-wave telescopes 
has been developed recently.\cite{GWA}  

In this paper, we propose alternative experiments 
possible with interferometers. 
Gravitational wave interferometers are 
huge macroscopic devices, 
yet, their measurement sensitivity is 
near the standard quantum limit 
by cutting-edge optical technology. 
Because of this high sensitivity, 
gravitational wave interferometers 
may be able to test Quantum Physics. 
Here, we focus on the quantum decoherence. 
As we discuss in detail below, 
the quantum decoherence is considered to be a key 
to solve the measurement problem of Quantum Physics. 
If we can measure that the quantum decoherence is occurring 
at a fundamental level, 
it will give an important clue 
to solving the measurement problem 
and will have major impact 
on the foundation of Quantum Physics. 

Because of the high sensitivity 
near the standard quantum limit, 
one may think that the quantization of the test mass 
is crucial to study the interferometer output. 
However, the test mass quantization is not relevant 
because its effect appears only in low frequency 
(around several {\it Hz})
and is normally filtered out from the data processing 
of the interferometer output.\cite{testmass} 
This limits the measurement possibility 
of the quantum decoherence acting on the test mass. 
Unless the quantum decoherence occurs quite frequently 
(such as by several hundreds {\it Hz}), 
one cannot read out its effect from the interferometer output. 
An idea was proposed 
to measure the decoherence effect on the test mass 
using the interferometer technology in Ref.\cite{deco}. 
The possibility to measure a certain decoherence mechanism 
acting on the test mass of the gravitational wave interferometers 
was also discussed in Ref.\cite{red,CSL}

Here, instead of the test mass, 
we consider the quantum decoherence of photons. 
There are a couple of advantages to measure 
the photon decoherence 
using the gravitational wave interferometers; 
\begin{itemize}
\item 
Gravitational wave interferometers 
are made to measure the photon interference 
and they can directly measure the photon decoherence. 
\item 
The photon is a fundamental particle 
and one can measure the fundamental decoherence. 
\item
For high sensitivity, 
the gravitational wave interferometers use high power laser. 
This will statistically increase the possibility 
of measuring decoherence. 
\item
The operation of the gravitational wave interferometers 
is very stable. 
For example, LIGO succeeded a year-long operation recently. 
This also helps statistically increasing the possibility 
of measuring decoherence. 
\item
As we discuss in Sec.\ref{sec:inf}, 
the decoherence measurement is shot-noise limited 
at relatively higher sideband frequency. 
In this region, many mechanical noise sources have already been 
identified and well-studied. 
This may serve a clean experimental setup. 
\item
The decoherence effect of the test mass can be seen 
as the noise of gravitational wave interferometers, 
therefore, it is hard to identify 
whether the decoherence actually occurs. 
On the other hand, as we show in Sec.\ref{sec:inf}, 
by changing the carrier light for the homodyne detection, 
one can distinguish the photon decoherence 
and the noise of the interferometers. 
\end{itemize}

The advanced design of gravitational wave interferometers 
uses the quantum non-demolition technology (hereafter QND)
to measure the photon interference.\cite{QND} 
In Sec.\ref{sec:deco} and Sec.\ref{sec:inf}, 
we discuss how the QND gravitational wave interferometers 
can measure the photon decoherence 
and its measurement limit. 
In Sec.\ref{sec:fun} and Sec.\ref{sec:grav}, 
we present some specific proposed models of 
the fundamental decoherence 
and discuss the measurement limit of the model parameters. 
In Sec.\ref{sec:con}, 
we conclude this paper.

%%%%%%%%%%%%%%%%%%%%%%%%%%%%%%%%%%%%%%%%%%%%%%%%%%%%%%%%%%%
\section{Interferometric Measurement of Quantum Decoherence}
\label{sec:deco}
%%%%%%%%%%%%%%%%%%%%%%%%%%%%%%%%%%%%%%%%%%%%%%%%%%%%%%%%%%%%

Gravitational wave interferometers 
have two arm cavities. 
Let us call two arm cavities by R-arm and L-arm. 
This huge device creates a superposition state 
from an injected photon, 
namely the superposition of the photon in the R-arm 
and the photon in the L-arm. 
By measuring the interference of these photon states, 
one is able to extract the information from gravitational waves. 

We denote the superposition state of the injected photon by 
\begin{eqnarray}
|\gamma_{in}>=\left(|\gamma_R>+|\gamma_L>\right)/\sqrt{2} 
\,, 
\end{eqnarray}
where $|\gamma_{R/L}>$ is the photon state in the R/L-arm. 
The interferometer has two output ports to be measured. 
One is called by the bright port and its state is described by 
\begin{eqnarray}
|\gamma_b>=\left(|\gamma_R> + |\gamma_L>\right)/\sqrt{2} \,. 
\end{eqnarray}
The other is the dark port which is 
\begin{eqnarray}
|\gamma_d>=\left(|\gamma_R> - |\gamma_L>\right)/\sqrt{2} \,. 
\end{eqnarray}
If the superposition state of the photon
has no external coupling, 
it is apparent that we will only see the bright port state. 
More specifically, the probability of measuring the photon 
coming out from the bright port 
is given as $P_b=\left|<\gamma_b|\gamma_{in}>\right|^2=1$ 
and the probability to measure the photon from the dark port 
is given as $P_d=\left|<\gamma_d|\gamma_{in}>\right|^2=0$. 

Let us assume 
that the superposition state of the photon 
is coupled to an external field $\phi$ 
while the photon state evolves in the interferometer. 
The photon in the R/L-arm induces 
the excitation of the external field 
and we can formally write the total state by 
\begin{eqnarray}
\left(|\gamma_R>\otimes |\phi_R> 
+ |\gamma_L>\otimes |\phi_L>\right)/\sqrt{2} \,, 
\end{eqnarray}
where the states $|\phi_{R/L}>$ describe the external fields 
that couples with the photon in the R/L-arm
and we ignore the back-action to the photon 
due to the field coupling. 
Taking the trace over the field state, 
the quantum state of the photon 
is described by the reduced density matrix as 
\begin{eqnarray}
\rho_\gamma &=& |\gamma_R><\gamma_R|/2 
+|\gamma_L><\gamma_L|/2 
\nonumber \\ 
&& +|\gamma_R><\gamma_L| \times <\phi_L|\phi_R>/2 
+|\gamma_L><\gamma_R| \times <\phi_R|\phi_L>/2 
\,, \label{eq:rden}
\end{eqnarray}
where we use 
$<\phi_R|\phi_R>=<\phi_L|\phi_L>=1$. 
When $Re\left[<\phi_R|\phi_L>\right]=0$, 
the interference term vanishes 
and the quantum state becomes effectively 
the classical sum over the R-state and the L-state. 
By this situation, we say 
the superposition state of the photon 
is decohered due to the field coupling. 

The probabilities to measure the photon 
coming from the bright port and from the dark port 
are given by 
\begin{eqnarray}
P_b &=& <\gamma_b|\rho_\gamma|\gamma_b> 
={1 \over 2} 
+{1 \over 2}Re\left[<\phi_R|\phi_L>\right]
\,, \\ 
P_d &=& <\gamma_d|\rho_\gamma|\gamma_d> 
={1 \over 2} 
-{1 \over 2}Re\left[<\phi_R|\phi_L>\right]
\,. 
\end{eqnarray} 
If $Re\left[<\phi_R|\phi_L>\right]=1$, 
the photon comes out of the bright port only. 
On the other hand, 
if $Re\left[<\phi_R|\phi_L>\right]=0$, 
the photon comes out of both the bright and dark ports 
with equal probability. 
Thus, the quantum decoherence of the superposition state 
can easily be observed 
by measuring the dark port of the interferometer.

%%%%%%%%%%%%%%%%%%%%%%%%%%%%%%%%%%%%%%%%%%%%%%%%%%%%%%%%%%%
\section{QND Laser Interferometer}
\label{sec:inf}
%%%%%%%%%%%%%%%%%%%%%%%%%%%%%%%%%%%%%%%%%%%%%%%%%%%%%%%%%%%%

In the previous section, we consider the ideal case 
that we inject a single photon into the interferometer 
and see the outcome of the dark port. 
In the actual operation of gravitational wave interferometers, 
we continuously inject the photon 
and the interferometer output is processed 
in the frequency domain. 
In this section, we discuss 
how we can see the photon decoherence 
and the measurement limit of the decoherence probability. 

We denote the injected light power by $I$. 
and the probability of the decoherence by $P_{deco}$. 
It is apparent that the decoherence of the photon state 
leads to the continuous light output from the dark port 
of the power $<I_{deco}> = I P_{deco}/2$ on average. 
However, one cannot measure 
this continuous component of the light output 
in the data analysis process of gravitational-wave detection 
because it uses the frequency decomposition 
of the light output with respect to the sideband $\Omega$ 
and the continuous component is filtered out of the data. 
Instead, we consider to measure the statistical fluctuation 
of the light output due to the decoherence. 

The number of photons injected to the interferometer 
in a short time interval $\Delta t$ 
is $N = I \Delta t/(\hbar \omega_0)$ 
where $\omega_0$ is the frequency of the carrier light. 
The statistical distribution of the number $n$ 
of the decohered photon state in this time interval 
follows the Poisson distribution 
$P[n] = e^{-P_{deco}N}{(P_{deco}N)^n \over n!}$. 
It is easy to see that the statistical average 
of the decohered photon state number is 
$<n> = \sum_n nP[n]  = P_{deco}N$ 
and the averaged light power from the dark port becomes 
\begin{eqnarray}
<I_{deco}> = {\hbar\omega_0 \over 2 \Delta t}<n> 
= I {P_{deco} \over 2} 
\,. 
\end{eqnarray} 
The statistical fluctuation of the decohered photon state number 
is $<(n-<n>)^2> = P_{deco}N$ 
and the fluctuation of the laser power 
we can measure from the dark port 
averaged over the time interval $\Delta t$ 
is derived as 
\begin{eqnarray}
<\Delta_{\Delta t} I_{deco}^2> 
= \left({\hbar\omega_0 \over 2 \Delta t}\right)^2 <(n-<n>)^2>
=I {\hbar\omega_0 \over \Delta t} {P_{deco} \over 4} 
\,. 
\end{eqnarray}
We suppose that the decoherence process is spontaneous 
and that decoherence events do not correlate with each other, 
in this case the output from the dark port becomes 
the white spectrum with respect to the sideband frequency. 
The two-time correlation of the light power fluctuation 
is given by 
\begin{eqnarray}
<\Delta I_{deco}(t)\Delta I_{deco}(t')> 
= {\hbar\omega_0 I P_{deco} \over 4} \times \delta(t-t') 
\,. 
\end{eqnarray}
The spectral density of the light power from the dark port 
$S_{I,deco}(\Omega)$ 
is the Fourier transformation of the two-time correlation 
defined by 
$<\Delta I_{deco}(t)\Delta I_{deco}(t')> = 
\int (d\Omega/2\pi) S_{I,deco}(\Omega) e^{i\Omega(t-t')} $ 
and we have 
\begin{eqnarray}
S_{I,deco}(\Omega) = {\hbar\omega_0 I P_{deco} \over 4}
\,. \label{eq:inf0}
\end{eqnarray}
If this is larger than the noise spectrum density, 
we may be able to observe the decoherence. 
In order to see the theoretical limit 
for the decoherence measurement 
and we consider an ideal QND interferometer 
without any mechanical noise. 

\bigskip
{\it 1) Decoherence Measurement by the QND interferometer} 

For gravitational wave detection, 
we use the so-called homodyne detection, 
where we measure a particular output quadrature field $E(t)$ 
by superposing the carrier light of the power $I_{carrier}$. 
The noise spectrum of the homodyne detection 
is given as (\ref{eq:ns_homo}) in Appendix. 
It is apparent that the noise spectrum becomes smaller 
for the smaller carrier light 
because the superposed carrier light amplifies the noise, 
but it does not amplify the decoherence signal. 
For the fixed carrier light power, 
one can lower the noise due to the opto-mechanical coupling 
by choosing the carrier light phase as $\alpha=0$ 
and we have the white noise spectrum as 
\begin{eqnarray}
S_{I,homo}(\Omega) &=& {\hbar \omega_0 I_{carrier} \over 4}
\,. \label{eq:inf1} 
\end{eqnarray}
This is because the quadrature field we measure 
is purely the amplitude quadrature 
which does not couple to the test mass location. 
Therefore, the noise does not include the back-action noise 
and it has only the shot noise 
which is a spontaneous process. 

Since the decoherence signal (\ref{eq:inf0}) 
and the noise (\ref{eq:inf1}) 
have the same spectrum shape, 
one cannot distinguish the decoherence signal from the noise 
for a fixed carrier light power. 
By changing the carrier light power, 
it is easy to extract the decoherence signal 
and that is the unique advantage 
in measuring the photon decoherence. 

The decoherence measurement is the shot-noise limited 
and its theoretical limit is simply given by 
\begin{eqnarray}
P_{deco} &>& {I_{carrier} \over I} 
\,. 
\end{eqnarray}
With the current LIGO operation, 
the noise curve reaches the fraction of the theoretical noise. 
The carrier light superposed onto the output quadrature field 
is around $I_{carrier} \approx 0.001 I$, 
therefore, even the successful operation 
of the the current gravitational wave interferometer 
constraints the decoherence probability by 
\begin{eqnarray}
P_{deco} \leq 0.001 \,. \label{eq:dm1} 
\end{eqnarray}

\bigskip
{\it 2) zero-carrier light} 

Because the homodyne detection amplifies 
the noise of the interferometer 
it is more advantageous for a decoherence measurement 
to directly measure the output light power by the photodetector 
without superposing the carrier light. 

The noise spectrum of the direct photo detection 
is given by (\ref{eq:ns_dir}) in Appendix. 
For simplicity, we consider the theoretical limit 
at specific sideband frequencies, $\Omega = 0, \gamma$, 
and the frequency, 
$\gamma << \Omega << (\omega_0^2 \gamma)^{1/3}$.
The theoretical limit is derived as 
\begin{eqnarray} 
P_{deco} &>& \left\{\matrix{
{\hbar \omega_0 \gamma \over I_{SQL}}
\left(10{I \over I_{SQL}}
+{429 \over 4}\left({I \over I_{SQL}}\right)^3\right) 
& \hbox{when $\Omega =0$,} \cr 
{\hbar \omega_0 \gamma \over I_{SQL}}
\left({29 \over 5}{I \over I_{SQL}}
+{29 \over 125}\left({I \over I_{SQL}}\right)^3\right) 
& \hbox{when $\Omega =\gamma$,}  \cr 
5{\hbar \omega_0 \gamma \over I_{SQL}}
{I \over I_{SQL}}
& \hbox{when $\gamma << \Omega << (\omega_0^2\gamma)^{1/3}$.} 
} \right. 
\,, 
\end{eqnarray}
which suggests that high-frequency data 
give us the better theoretical limit 
for the decoherence measurement. 
It is interesting to point out 
that the noise has stronger dependence 
on the injected light power $I$ 
and that we may not be able to measure the decoherence 
if the injected light is too strong. 

Suppose we inject the SQL light power 
to a LIGO-scale interferometer, 
$I = I_{SQL}= 1.0 \times 10^4 W = 1.0 \times 10^{11} erg/sec$, 
$\omega_0 = 1.8 \times 10^{15} sec^{-1}$ 
and $\gamma = 2\pi \times 100 sec^{-1}$, 
we have the theoretical limit 
for the decoherence measurement as 
\begin{eqnarray} 
P_{deco} > 6.0 \times 10^{-20} \,. \label{eq:dm2} 
\end{eqnarray}

%%%%%%%%%%%%%%%%%%%%%%%%%%%%%%%%%%%%%%%%%%%%%%%%%%%%%%%%%%%
\section{Decoherence by Spontaneous Localization}
\label{sec:fun}
%%%%%%%%%%%%%%%%%%%%%%%%%%%%%%%%%%%%%%%%%%%%%%%%%%%%%%%%%%%%

Quantum Physics is based on two distinct processes. 
One is the unitary evolution process of the state, 
which is linear, deterministic and reversible. 
The other is according to {\it Copenhagen interpretation} 
the state reduction process 
at the time of measuring the quantum state, 
which is non-linear, stochastic and irreversible. 
These processes are incompatible 
and the identification of the borderline 
between the regimes these two processes to be applied 
is called by {\it the measurement problem}. 
Because quantum measurement involves 
a macroscopic measurement device 
with a huge number of degrees of freedom 
interacting with surrounding environment, 
it is far from trivial to identify 
when the state reduction process would take over 
the dynamical process of Quantum Physics. 

The paradoxical nature of this measurement problem 
is best illustrated 
by the famous Schr\"odinger's cat;\cite{cat} 
a cat in a box is evolving into linear superposition 
of the quantum state where the cat is `alive' 
and the state where the cat is `dead'. 
Now we open the box and measure the cat. 
Suddenly the superposition state disappears 
and we will see the cat is either alive or dead 
with nonzero classical probability. 
We then ask; 
Why do we see only the state of the living cat 
or the state of the dead cat? 
Why don't we see the superposition state 
of the living cat and the dead cat? 
What dynamical process transforms 
the quantum superposition 
into the classical sum of these specific states? 

One of proposals solving this issue is 
to unify these two physical processes 
by adding small non-linear and stochastic effects 
to the unitary evolution process. 
This new stochastic effects are made to be small 
for a microscopic system 
so that we still have the unitary evolution, 
but, the effect becomes large for a macroscopic system. 
When a system is measured by a macroscopic measurement device, 
the stochastic effects strongly act on the device 
and the state reduction process occurs dynamically 
to the quantum state of the device. 

Along this idea, several models have been proposed 
to modify the Schr\"odinger equations 
which govern the unitary evolution process of the quantum state. 
One of the simplest models is called 
Quantum Mechanics with Spontaneous Localization 
(hereafter QMSL). 
In this model, we specifically consider 
the quantum evolution of a collection of particles. 
Most of the time, the wave-function evolves 
by the usual Schr\"odinger equations, 
but, spontaneously at a mean rate $\lambda$, 
the localization process occurs 
following a probability distribution derived from the wave-function. 
By the localization process, the localization operator 
is applied to the wave-function 
which localizes the wave packet of each particle 
by the length scale $1/\sqrt{\alpha}$. 
This process collapses the wave-function dynamically 
in the position-representation without measurement. 
An interesting feature of QMSL is that, 
although the localization operator localizes 
the wave packet of each particle by the rate $\lambda$, 
it cumulatively localizes 
the reduced density matrix of the center of mass 
by the rate $N \lambda$ 
where $N$is the number of the particles. 
This means that, for a microscopic system, 
the localization effect becomes negligible 
with sufficiently small $\lambda$, 
but, it could be substantially large for a macroscopic system 
because of extremely large $N$. 
According to Ref.\cite{reduction}, it was suggested 
that the model parameters could be 
\begin{eqnarray}
\lambda \simeq 10^{-16} [s^{-1}]  \,, \quad 
1/\sqrt{\alpha} \simeq 10^{-5} [cm] \,. 
\label{eq:QMSL} 
\end{eqnarray} 
More thorough survey of various experiments\cite{CSL} 
indicates that 
\begin{eqnarray}
\lambda \simeq 10^{-7 \pm 2} [s^{-1}]  \,, \quad 
1/\sqrt{\alpha} \simeq 10^{-5} [cm] \,. 
\label{eq:CSL} 
\end{eqnarray}

Let us suppose that the photon dynamics has 
a similar spontaneous localization process. 
Gravitational wave interferomters are sufficiently large 
compared to the localization scale $1/\sqrt{\alpha}$, 
thus, once the spontaneous localization occurs 
to the photon propagating the detectors, 
the photon state becomes the classical sum 
of the photon state in the R-arm 
and the photon state in the L-arm, 
that is, the photon state becomes completely decohered. 
As we discussed in Sec.\ref{sec:deco}, 
this can be measured by the photo-detection at the dark port. 
The decoherence probability is 
the mean rate of the spontaneous localization 
multiplied by the photon propagation time as 
\begin{eqnarray}
P_{deco} = \lambda/\gamma \,. 
\end{eqnarray} 
For a LIGO-scale interferometer 
with the suggested parameters of QMSL (\ref{eq:QMSL}), 
we have the decoherence probability as 
\begin{eqnarray}
P_{deco} \sim 10^{-19}  \,. 
\end{eqnarray} 
If we use the parameters (\ref{eq:CSL}), 
the decoherence probability would be 
\begin{eqnarray}
P_{deco} \sim 10^{-10 \pm 2}  \,. 
\end{eqnarray} 

These values are consistent with the fact 
that the current gravitational wave detectors do not 
see the decoherence effect by the homodyne detection 
since it satisfies (\ref{eq:dm1}). 
However, if one can prepare 
the ideal dark port with the zero-carrier light, 
one may be possibly measure the decoherence 
due to the spontaneous localization 
according to (\ref{eq:dm2}).

%%%%%%%%%%%%%%%%%%%%%%%%%%%%%%%%%%%%%%%%%%%%%%%%%%%%%%%%%%%
\section{Gravitational decoherence - semi-classical gravity}
\label{sec:grav}
%%%%%%%%%%%%%%%%%%%%%%%%%%%%%%%%%%%%%%%%%%%%%%%%%%%%%%%%%%%%

It was suggested that gravity might serve 
the mechanism for the fundamental decoherence.\cite{grav} 
In this section, we estimate the decoherence probability 
due to the gravitational coupling of the interferometer 
using the semi-classical gravity. 

Let us denote the probability of the graviton emission by $P_g$ 
for the photon in either the R-arm or in the L-arm. 
By the interferometer, 
we have the photon superposition state, 
$(|\gamma_R> +|\gamma_L>)/\sqrt{2}$. 
Due to the gravitational coupling, 
the photon state evolves into 
\begin{eqnarray}
|\gamma_R>/\sqrt{2}\otimes(|0>+c_R|g_R>)/\sqrt{1+P_g}
+|\gamma_L>/\sqrt{2}\otimes(|0>+c_L|g_L>)/\sqrt{1+P_g}
\,, 
\end{eqnarray}
where $|0>$ is the zero-graviton state 
and $|g_R>$ and $|g_L>$ are the one-graviton state 
emitted from the photon in the R-arm and the L-arm 
of the interferometer. 
For simplicity, we here ignore the back-action to the photon 
due to the graviton emission. 
$c_R$ and $c_L$ are the amplitudes of the graviton emission 
and we have $|c_R|^2=|c_L|^2=P_g$. 
We suppose that the scale of the interferometer is large enough 
so that there is no interference 
between the one-graviton states 
emitted from the R-arm and the L-arm, 
i.e. $<g_R|g_L> \approx 0$. 
Then the reduced density matrix for the photon state 
can be written as 
\begin{eqnarray}
{1 \over 2}(|\gamma_R><\gamma_R|+|\gamma_L><\gamma_L|) 
\nonumber \\ 
+{1 \over 2(1+P_g)}
(|\gamma_R><\gamma_L|+|\gamma_L><\gamma_R|)
\,. 
\end{eqnarray}
This shows that 
the decoherence probability of the photon state 
is approximately equal to the graviton emission probability, 
$P_{deco} \simeq P_g$. 

In the following, we calculate the emission probability 
using the semi-classical gravity. 
We are especially interested in the observational possibility 
to distinct some gravitational models 
and we consider the standard $4$-D gravity, 
and the brane gravity. 

\bigskip
{\it Standard $4$-D gravity}

Photons that propagate freely do not emit gravitons. 
When the photon hits the mirrors of the arm cavity, 
the orbit of the photon moves from a null geodesic to another 
by exchange its momentum with the mirror, 
and it emits gravitons. 

Although the photon is a relativistic particle, 
we use the quadrupole formula to estimate 
the emission probability of graviton. 
The energy flux is given by 
\begin{eqnarray}
L = {G \over 5 c^5} \left|{d^3 \hat{\bf I}\over dt^3}\right|^2 \,, 
\end{eqnarray}
where $\hat{\bf I}$ is the trace-free component 
of the mass quadrupole moment tensor of the system. 
When the photon hits the mirror, 
they exchange the momentum 
and produce the nontrivial quadrupole. 
This process occurs in the mirror's surface layer 
which has the thickness of the photon wavelength 
in the time scale of the photon frequency. 
Thus, we assume $d^3 {\bf I}/dt^3 \approx \hbar \omega_0^2$ 
and that the duration of the graviton emission 
is $\approx 1/\omega_0$. 
Diving the energy flux by the graviton energy 
($\approx \hbar \omega_0$), 
the number of the emitted gravitons is estimated as 
\begin{eqnarray}
N \approx {G\hbar \omega_0^2 \over c^5} \,. 
\end{eqnarray}
We regard $N$ as the probability of the graviton emission 
because $N<<1$. 
The emission probability is multiplied by the finesse $1/T$ 
because the photon is trapped by the highly reflective mirrors 
of the cavity 
($T$ is the mirrors' transmissivity of the cavity.)
The total probability of the decoherence is obtained as 
\begin{eqnarray}
P_{deco} \approx {G\hbar \omega_0^2 \over c^5 T} 
\,. \label{eq:g-deco}
\end{eqnarray}
For the LIGO-scale interferometer, this becomes 
\begin{eqnarray}
P_{deco} \approx 10^{-55} \,, 
\end{eqnarray}
thus, it is unlikely to measure the gravitational decoherence.

\bigskip
{\it Brane gravity}

Brane cosmology was suggested 
as an interesting solution of the hierarchical problem 
in particle physics. 
In the brane cosmology, 
spacetime is $(4+1)$-dimensional 
but, matters are located 
only on a $(3+1)$-hypersurface ($3$-brane). 
Because of the extra-dimensional freedom 
of the gravitational field, 
it has an extra channel to emit gravitons, 
which is called by the Kaluza-Klein mode (hereafter, KK-mode). 
The size of the extra-dimension ($\ell$) is not well-constrained 
by the present experiments of gravitational forces 
if it is smaller than $0.1mm$.\cite{extra} 
If the characteristic wavelength of graviton emission 
is smaller than the size of the extra-dimension, 
we may expect huge emission of KK-mode gravitons. 

The modified quadrupole formula was derived\cite{quad} 
and the energy flux through the KK-modes is given as 
\begin{eqnarray}
L_{KK} = -{ell^2 \over 18 c^2} {d^5 I \over dt^5} 
\,, \label{eq:KKg}
\end{eqnarray}
where $I$ is the trace of the mass quadrupole. 
The probability of the KK-mode graviton emission
%\footnote{Because (\ref{eq:KKg}) is not positive definite,  
%it could be the case that there is no KK-mode graviton emission, 
%instead, the photon absorbs 
%the geometrical energy of the background curvature. 
%For example, we suppose that the orbit of the photon 
%is described as 
%$x=(c/\omega_0)\cosh(u), t=(1/\omega_0)\sinh(u)$. 
%In this case, we have $d^5 x/dt^5 = 
%}
is estimted as 
\begin{eqnarray}
P_{KK-deco} \approx {\ell^2 \omega_0^2 \over c^2 T} 
\,. \label{eq:k-deco}
\end{eqnarray}
For the LIGO-scale interferometer, this becomes 
\begin{eqnarray}
P_{KK-deco} \approx \left({\ell \over 0.1 \hbox{mm}}\right)^2 10^7 
\,. 
\end{eqnarray}
This successful operation of the current LIGO implies 
that $P_{KK-deco} < 0.001$, 
which constrains the size of the extra-dimension to 
\begin{eqnarray}
\ell < 10^{-6} \hbox{mm} \,. 
\end{eqnarray} 
With the zero-carrier light limit, 
the gravitational wave detectors can measure 
the gravitational decoherence 
due to the KK-mode graviton emission if 
\begin{eqnarray}
\ell > 10^{-14} \hbox{mm} \,. 
\end{eqnarray}

%%%%%%%%%%%%%%%%%%%%%%%%%%%%%%%%%%%%%%%%%%%%%%%%%%%%%%%%%%%
\section{Conclusion}
\label{sec:con}
%%%%%%%%%%%%%%%%%%%%%%%%%%%%%%%%%%%%%%%%%%%%%%%%%%%%%%%%%%%%

The difficulty in measuring a specific decoherence process 
is that there is not way to separate 
the decoherence signal and the noise from the measurement output. 
Both the decoherence and the noise usually appear 
in the measurement output in the quite similar way 
because they come from the system's coupling 
with uncontrolled environment. 
As we see from Sec.\ref{sec:inf}, 
this is actually the case 
when we consider the decoherence measurement 
using interferometers, 
i.e. the decoherence signal and the quantum noise 
have the common spectrum shape. 
However, the quantum noise is proportional 
to the light power of the carrier light, 
thus, it is possible to separate 
the noise and the decoherence signal 
by changing the carrier light. 
This allow us to estimate the measurement limit 
of the decoherence probability 
and we obtain the following results: 
\begin{itemize}
\item
The current successful operation of LIGO shows that 
the decoherence probability must be 
\begin{eqnarray}
P_{deco} \leq 0.001 \,.
\end{eqnarray}
\item
By taking the zero carrier-light limit, 
LIGO-scale interferometers can measure the decoherence 
if the decoherence probability is 
\begin{eqnarray}
P_{deco} > 6.0 \times 10^{-20} \,. 
\end{eqnarray}
\end{itemize}

We consider two kinds of the decoherence mechanism. 
One is the spontaneous localization 
as the modification of the Schr\"odinger equation 
and it has two model parameters. 
The proposed value of these parameters are 
just consistent with the current operation of LIGO, 
but, if we could take the zero carrier-light limit, 
it is possible to see the fundamental decoherence 
due to this mechanism, 
otherwise, we could experimentally deny 
this modification of Quantum Physics. 

We also consider gravitational decoherence. 
For standard $4$D semi-classical gravity, 
it is almost impossible to see 
the gravitational decoherence of photons. 
If we consider a extra-dimensional gravity model, 
one could see the strong decoherence 
due to the coupling with the bulk gravitational field. 
We show that the brane world gravity is such a case 
and we find that the size of the extra-dimension 
must be constrained by 
\begin{eqnarray}
\ell < 10^{-6} \,, 
\end{eqnarray} 
by the current operation of the LIGO experiment.

%%%%%%%%%%%%%%%%%%%%%%%%%%%%%%%%%%%%%%%%%%%%%%%%%%%%%%%%%%%
%%%%%%%%%%%%%%%%%%%%%%%%%%%%%%%%%%%%%%%%%%%%%%%%%%%%%%%%%%%

\section{Acknowledgment}
We thank Prof. Kip Thorne, Prof. Yanbei Chen 
and Dr. Jeandrew Brink for fruitful discussion. 
This work is supported 
by NSF grant PHY-0601459, PHY-0653653, 
NASA grant NNX07AH06G, NNG04GK98G 
and the Brinson Foundation.

%%%%%%%%%%%%%%%%%%%%%%%%%%%%%%%%%%%%%%%%%%%%%%%%%%%%%%%%%%%
%%%%%%%%%%%%%%%%%%%%%%%%%%%%%%%%%%%%%%%%%%%%%%%%%%%%%%%%%%%

\appendix

%%%%%%%%%%%%%%%%%%%%%%%%%%%%%%%%%%%%%%%%%%%%%%%%%%%%%%%%%%%
\section{Quantum description of QND Laser Interferometer}
\label{app:qnd}
%%%%%%%%%%%%%%%%%%%%%%%%%%%%%%%%%%%%%%%%%%%%%%%%%%%%%%%%%%%%

The standard (single-mode) expression 
of the quantized $1$-dimensional optical field 
is given as 
\begin{eqnarray}
E &=& \int^\infty_0 {d\omega \over 2\pi} 
\sqrt{4\pi \hbar \omega \over A c}
\left(a e^{-i\omega t} +a^\dagger e^{i\omega t}\right) 
\,, \label{eq:one-p} 
\end{eqnarray}
where $A$ is the cross section area of the beam. 
$a$ and $a^\dagger$ are the annihilation and creation operators 
that satisfy the commutation 
\begin{eqnarray}
[a(\omega),a(\omega')]=0 \,, \quad 
[a(\omega),a^\dagger(\omega')]=2\pi\delta(\omega-\omega') 
\,. 
\end{eqnarray}
We are interested in the optical fields 
of the sideband $\omega = \omega_0+\Omega$ 
about the carrier light frequency $\omega_0$. 
In this case, the optical field is conveniently described 
by the two-photon formalism as 
\begin{eqnarray}
E &=& E^{(c)}\cos(\omega_0 t) +E^{(s)}\sin(\omega_0 t)
\,, \\ 
E^{(c)} &=& \sqrt{4\pi \hbar \omega_0 \over A c}
\int^\infty_0 {d\Omega \over 2\pi}
\left(a^{(c)} e^{-i\Omega t} +a^{(c)\dagger} e^{i\Omega t}\right) 
\,, \\ 
E^{(s)} &=& \sqrt{4\pi \hbar \omega_0 \over A c}
\int^\infty_0 {d\Omega \over 2\pi}
\left(a^{(s)} e^{-i\Omega t} +a^{(s)\dagger} e^{i\Omega t}\right) 
\,, 
\end{eqnarray} 
where $\Omega$ is the sideband frequency 
and we only consider the domain where $\Omega << \omega_0$. 
$E^{(c)}$ and $E_s$ 
are called by the cosine and sine quadrature fields. 
$a^{(c)}(\Omega)$ and $a^{(s)}(\Omega)$ are the quantum operators 
and are related to the creation/annihilation operators 
of the single-photon mode by 
\begin{eqnarray}
a^{(c)}(\Omega) = {1 \over \sqrt{2}}
\left(a(\omega_0+\Omega)+a^\dagger(\omega_0-\Omega)\right) 
\,, \quad 
a^{(s)}(\Omega) = {1 \over i\sqrt{2}}
\left(a(\omega_0+\Omega)-a^\dagger(\omega_0-\Omega)\right) 
\,. 
\end{eqnarray} 

The interferometer is an optical device 
to produce the poderomotive squeezing of the optical fields 
through its arm cavities. 
With the general expression of the optical field 
in the two-photon formalism, 
we use the quantum operators $a^{(c)}_{in}$ and $a^{(s)}_{in}$ 
in the place of $a^{(c)}$ and $a^{(s)}$ for the input optical field 
and we use the quantum operators $a^{(c)}_{out}$ and $a^{(s)}_{out}$ 
for the output optical field. 
For an ideal interferometer, these operators are related by 
\begin{eqnarray}
a^{(c)}_{out} = a^{(c)}_{in} e^{2i\beta} 
\,, \quad 
a^{(s)}_{out} = \left(a^{(s)}_{in}-{\cal K}a^{(c)}_{in}\right)e^{2i\beta} 
+\sqrt{2{\cal K}}{h \over h_{SQL}}e^{i\beta} 
\,, \label{eq:in-out} 
\end{eqnarray} 
where we use 
$\beta = \arctan(\Omega/\gamma)$ 
as the phase shift of off-resonant optical fields, 
${\cal K} = 2(I/I_{SQL})\gamma^4/\Omega^2/(\Omega^2+\gamma^2)$ 
as the opto-mechanical coupling constant, 
$I_{SQL} = mL^4\gamma^4/(4\omega_0)$ 
as the laser light power to reach the standard quantum limit, 
and $h_{SQL} = \sqrt{8\hbar/(m\Omega^2L^2)}$ 
as the standard quantum limit 
for the gravitational-wave measurement. 
($\gamma$ is the cavities' half bandwidths, 
$m$ is the reduced test mass of the interferometer 
and $L$ is the arm length.) 
$h$ is the Fourier transform of the gravitational-wave signal $h(t)$ 
defined by 
\begin{eqnarray}
h = \int^\infty_{-\infty} dt e^{i\Omega t} h(t) 
\,, 
\end{eqnarray} 
and the gravitational-wave signal can be read out 
from the sine quadrature of the output field. 

\bigskip

{\it 1) Homodyne detection}

By the homodyne detection, 
it is possible to measure the output field 
by the direct photo detection. 
With the beam splitter, 
we superpose the laser light of the carrier frequency $\omega_0$ 
on the output field 
\begin{eqnarray} 
E &=& \sqrt{4\pi \hbar \omega_0 \over A c}
\sqrt{2}D\cos(\omega_0 t+\alpha) 
\,, 
\end{eqnarray}
where the light power is $\hbar \omega_0 D^2$. 
The resulting optical field for the homodyne detection 
is written as 
\begin{eqnarray}
E_{homo} &=& E^{(c)}_{homo}\cos(\omega_0 t) 
+E^{(s)}_{homo}\sin(\omega_0 t)
\,, \\ 
E^{(c)}_{homo} &=& \sqrt{4\pi \hbar \omega_0 \over A c}
\left[D\cos(\alpha)
+{1 \over \sqrt{2}}\int^\infty_0 {d\Omega \over 2\pi}
\left(a^{(c)}_{out} e^{-i\Omega t} +a^{(c)}_{out}{}^\dagger e^{i\Omega t}\right)
\right]
\,, \\ 
E^{(s)}_{homo} &=& \sqrt{4\pi \hbar \omega_0 \over A c}
\left[-D\sin(\alpha)
+{1 \over\sqrt{2}}\int^\infty_0 {d\Omega \over 2\pi}
\left(a^{(s)}_{out} e^{-i\Omega t} +a^{(s)}_{out}{}^\dagger e^{i\Omega t}\right) 
\right]
\,. 
\end{eqnarray} 
The light power to be detected by the photodetector becomes 
\begin{eqnarray}
I_{homo}(t) &=& {E^2_{homo} \over 4\pi}Ac 
\nonumber \\ 
&\approx& \hbar \omega_0 \left[D^2 
+{D \cos(\alpha) \over \sqrt{2}}
\int^\infty_0 {d\Omega \over 2\pi}
\left(a^{(c)}_{out} e^{-i\Omega t} +a^{(c)}_{out}{}^\dagger e^{i\Omega t}\right)
-{D \sin(\alpha) \over \sqrt{2}}
\int^\infty_0 {d\Omega \over 2\pi}
\left(a^{(s)}_{out} e^{-i\Omega t} +a^{(s)}_{out}{}^\dagger e^{i\Omega t}\right)
\right]
\,, 
\end{eqnarray} 
where we ignore the very high frequency mode. 

For the conventional interferometer, 
the input quantum state is the vacuum state 
with respect to the in-state quantum operator, 
i.e. $a(\omega)|0>=0$. 
Using (\ref{eq:in-out}), 
we obtain the expectation value of the detected light power, 
\begin{eqnarray}
<I_{homo}>(t) &=& \hbar \omega_0 \left[D^2 
-D \sin(\alpha)\int^\infty_{-\infty} {d\Omega \over 2\pi}
\sqrt{\cal K}{h \over h_{SQL}}e^{i\beta-i\Omega t}
\right]
\,, \\ 
<I_{homo}>(\Omega) &:=& \int dt e^{i\Omega t} <I_{homo}>(t) 
= -D \sin(\alpha) \hbar \omega_0 
\sqrt{\cal K}{h \over h_{SQL}}e^{i\beta}
\,. 
\end{eqnarray} 
The quantum fluctuation of the light power 
$\delta I_{homo} = I_{homo}-<I_{homo}>$ 
becomes 
\begin{eqnarray}
<\delta I_{homo}(t)\delta I_{homo}(t')> &=& 
\int {d\Omega \over 2\pi} S_{I,homo}(\Omega) 
e^{-i\Omega(t-t')}
\,, \\ 
S_{I,homo}(\Omega) &=& {(\hbar \omega_0)^2 \over 4}D^2
\left\{ \left(\cos(\alpha)+{\cal K}\sin(\alpha)\right)^2
+\sin^2(\alpha)\right\}
\,, \label{eq:ns_homo}
\end{eqnarray}
where we use 
\begin{eqnarray} 
&& <a^{(c)}_{out}(\Omega)a^{(c)}_{out}{}^\dagger(\Omega')> 
= <a^{(c)}_{out}{}^\dagger(\Omega)a^{(c)}_{out}(\Omega')> 
= \pi \delta(\Omega-\Omega')
\,, \\ 
&& <a^{(c)}_{out}(\Omega)a^{(s)}_{out}{}^\dagger(\Omega')> 
= <a^{(c)}_{out}{}^\dagger(\Omega)a^{(s)}_{out}(\Omega')> 
= \pi (i -{\cal K})  \delta(\Omega-\Omega')
\,, \\ 
&& <a^{(s)}_{out}(\Omega)a^{(c)}_{out}{}^\dagger(\Omega')> 
= <a^{(s)}_{out}{}^\dagger(\Omega)a^{(c)}_{out}(\Omega')> 
= \pi (-i -{\cal K})  \delta(\Omega-\Omega')
\,, \\ 
&& <a^{(s)}_{out}(\Omega)a^{(s)}_{out}{}^\dagger(\Omega')> 
= <a^{(s)}_{out}{}^\dagger(\Omega)a^{(s)}_{out}(\Omega')> 
= \pi (1 +{\cal K}^2)  \delta(\Omega-\Omega')
\,. 
\end{eqnarray}

\bigskip

{\it 2) Direct detection}

The homodyne detection is advantageous 
for the gravitational wave measurement 
because the signal is amplified by the carrier light. 
However, that is not the case 
for the decoherence measurement 
because the quantum noise is also amplified. 
We therefore consider the detection 
at the zero-carrier-light limit. 
In this case, the light power 
to be detected by the photodetector becomes 
\begin{eqnarray}
I_{dir}(t) &\approx& {\hbar\omega_0 \over 2}\Biggl[
\left\{\int {d\Omega \over 2\pi}
\left(a^{(c)}_{out} e^{-i\Omega t} +a^{(c)}_{out}{}^\dagger e^{i\Omega t}\right)
\right\}^2
+\left\{\int^\infty_0 {d\Omega \over 2\pi}
\left(a^{(s)}_{out} e^{-i\Omega t} +a^{(s)}_{out}{}^\dagger e^{i\Omega t}\right)
\right\}^2
\Biggl] 
%I_{dir}(t) &=& {\hbar\omega_0 \over 2}\Biggl[
%\left(1+\cos(2\omega_0t)\right)
%\left\{\int {d\Omega \over 2\pi}
%\left(a^{(c)}_{out} e^{-i\Omega t} +a^{(c)}_{out}{}^\dagger e^{i\Omega t}\right)
%\right\}^2
%+\left(1-\cos(2\omega_0t)\right)
%\left\{\int^\infty_0 {d\Omega \over 2\pi}
%\left(a^{(s)}_{out} e^{-i\Omega t} +a^{(s)}_{out}{}^\dagger e^{i\Omega t}\right)
%\right\}^2
%\nonumber \\ && \qquad 
%+\sin(2\omega_0t)
%\biggl\{\int {d\Omega \over 2\pi}
%\left(a^{(c)}_{out} e^{-i\Omega t} +a^{(c)}_{out}{}^\dagger e^{i\Omega t}\right)
%\int^\infty_0 {d\Omega' \over 2\pi}
%\left(a^{(s)}_{out} e^{-i\Omega' t} +a^{(s)}_{out}{}^\dagger e^{i\Omega' t}\right)
%\nonumber \\ && \qquad \qquad \qquad \qquad 
%+\int {d\Omega \over 2\pi}
%\left(a^{(s)}_{out} e^{-i\Omega t} +a^{(s)}_{out}{}^\dagger e^{i\Omega t}\right)
%\int^\infty_0 {d\Omega' \over 2\pi}
%\left(a^{(c)}_{out} e^{-i\Omega' t} +a^{(c)}_{out}{}^\dagger e^{i\Omega' t}\right)
%\biggr\} \Biggr]
\,. 
\end{eqnarray} 

The expectational value of the light power 
with the vacuum input state is obtained as 
\begin{eqnarray}
<I_{dir}(t)> &\approx& \hbar\omega_0
\left\{\int^\infty_{-\infty} {d\Omega \over 2\pi}
\sqrt{\cal K}{h \over h_{SQL}}e^{i\beta}e^{-i\Omega t}
\right\}^2
+{\hbar\omega_0 \over 4}
\int {d\Omega \over 2\pi}{\cal K}^2
\nonumber \\ 
&=& \hbar\omega_0
\left\{\int^\infty_{-\infty} {d\Omega \over 2\pi}
\sqrt{\cal K}{h \over h_{SQL}}e^{i\beta}e^{-i\Omega t}
\right\}^2
+{5 \over 4}\hbar\omega_0\left({I \over I_{SQL}}\right)^2\gamma
\,. 
\end{eqnarray} 
The second term is due to the poderomotive squeezing 
of the interferometer 
and we subtract the vacuum contribution for regularization. 
The integration of ${\cal K}^2$ over $\Omega$ 
becomes infinite for small $|\Omega|$ 
since ${\cal K} \propto 1/\Omega^2$. 
This low-frequency behavior 
comes from the free motion of the mirror. 
We regulate the divergence by using 
${\cal K} = 2(I/I_{SQL})\gamma^4
/(\Omega-i\epsilon)^2/(\Omega^2+\gamma^2)$, 
which corresponds 
to adding an infinitesimally small damping force 
to the mirror motion. 

The quantum fluctuation of the light power is obtained as 
\begin{eqnarray}
<\{\delta I_{dir}(t),\delta I_{dir}(t')\}> 
&\approx& {(\hbar\omega_0)^2 \over 8}
\int^\infty_{-\infty}{d\Omega \over 2\pi}
\int^\infty_{-\infty}{d\Omega' \over 2\pi}
\Biggl\{1
+\left(i-{\cal K}(\Omega)\right)\left(i-{\cal K}(\Omega')\right)
+\left(-i-{\cal K}(\Omega)\right)\left(-i-{\cal K}(\Omega')\right)
\nonumber \\ && \qquad \qquad \qquad \qquad \qquad \qquad 
+\left(1+{\cal K}^2(\Omega)\right)\left(1+{\cal K}^2(\Omega')\right)
\Biggr\}e^{-i(\Omega+\Omega')(t-t')}
\nonumber \\ 
%&=& {(\hbar\omega_0)^2 \over 8}
%\int^\infty_{-\infty}{d\Omega \over 2\pi}
%int^\infty_{-\infty}{d\Omega' \over 2\pi}
%\left\{\left({\cal K}(\Omega)+{\cal K}(\Omega')\right)^2
%+{\cal K}^2(\Omega){\cal K}^2(\Omega')\right\}
%e^{-i(\Omega+\Omega')(t-t')}
%\nonumber \\ 
&=& \int^\infty_{-\infty} {d\Omega \over 2\pi} 
S_{I,dir}(\Omega) e^{-i\Omega(t-t')}
\,, \\ 
S_{I,dir}(\Omega) &\approx& {(\hbar \omega_0)^2 \over 16}
\int^\infty_{-\infty}{d\bar\Omega \over 2\pi}
\left\{\left({\cal K}(\Omega_+)+{\cal K}(\Omega_-)\right)^2
+{\cal K}^2(\Omega_+){\cal K}^2(\Omega_-)
\right\}_{\Omega_\pm = (\Omega\pm\bar\Omega)/2}
\nonumber \\ 
&=& (\hbar \omega_0)^2 \Biggl\{
{1 \over 4}\left({I \over I_{SQL}}\right)^2 
\left\{ 5\gamma 
+{4 \gamma^5(5\gamma^2-\Omega^2) \over 
(\Omega^2+\gamma^2)^2 (\Omega^2+4\gamma^2)}
\right\}
\nonumber \\ && \qquad \quad 
+16 \left({I \over I_{SQL}}\right)^4 
{\gamma^9(1716\gamma^8-715\gamma^6\Omega^2
-741\gamma^4\Omega^4-33\gamma^2\Omega^6
+5\Omega^8) \over 
(\Omega^2+\gamma^2)^5(\Omega^2+4\gamma^2)^3} \Biggr\}
\,. \label{eq:ns_dir}
\end{eqnarray} 
We note that the $\Omega$-integration here is consistent 
with the approximation of the two-photon formalism, 
$|\Omega|<\omega_0$ 
because ${\cal K}$ becomes small for large $|\Omega|$. 

The noise spectrum (\ref{eq:ns_dir}) 
does not include the shot noise 
because it comes entirely from the opto-mechanical coupling. 
In the case of the homodyne detection, 
the quantum fluctuation of the output quadrature field 
around the carrier light frequency, 
$\omega \approx \omega_0$ 
is converted into the fluctuation at the sideband frequency 
by the superposition of the carrier light. 
On the other hand, there is no such conversion process 
for the direct detection. 
The shot noise comes from the fluctuation 
at the low frequency, $\omega \approx 0 << \omega_0$ 
in the one-photon formalism, 
which is not properly described by the two-photon formalism. 
Hence, it is reasonable that 
the noise spectrum (\ref{eq:ns_dir}) 
does not include the shot noise. 

The shot noise to be added to (\ref{eq:ns_dir}) 
can be derived from the one-photon formalism. 
Because it comes from the quantum fluctuation 
of the input vacuum, 
we can use the standard form (\ref{eq:one-p}) 
and we have 
\begin{eqnarray}
<\{\delta I_{dir}(t),\delta I_{dir}(t')\}>_{shot} 
&=& \int^\infty_0 {d\omega \over 2\pi}
\int^\infty_0 {d\omega' \over 2\pi}
2\hbar^2(\omega\omega')
\cos\left((\omega+\omega')(t-t')\right)
\nonumber \\ 
&=& \int^\infty_{-\infty} {d\Omega \over 2\pi} 
S_{I,shot}(\Omega) e^{-i\Omega(t-t')}
\,, \\ 
S_{I,shot}(\Omega) 
%&=& \int^\infty_0 {d\omega \over 2\pi}
%\int^\infty_0 {d\omega' \over 2\pi}
%2\hbar^2(\omega\omega')
%2\pi\delta(\Omega-\omega-\omega') 
%\nonumber \\ 
&=& {\hbar^2 \Omega^3 \over 6\pi}
\,. \label{eq:ns_shot}
\end{eqnarray}
The shot noise is small 
for $\Omega < (\omega_0^2 \gamma)^{1/3}$ 
and we ignore this noise in Sec.\ref{sec:inf}.

%%%%%%%%%%%%%%%%%%%%%%%%%%%%%%%%%%%%%%%%%%%%%%%%%%%%%%%%%%%

%%%%%%%%%%%%%%%%%%%%%%%%%%%%%%%%%%%%%%%%%%%%%%%%%%%%%%%%%%%

\end{document}